\documentclass[conference]{IEEEtran}
\IEEEoverridecommandlockouts
\usepackage{cite}
\usepackage{amsmath,amssymb,amsfonts}
\usepackage{algorithmic}
\usepackage{graphicx}
\usepackage{textcomp}
\usepackage{booktabs}
\usepackage{xcolor}
\usepackage[most]{tcolorbox}
\tcbset{
  reviewbox/.style={
    enhanced,
    colback=gray!5,
    colframe=black, 
    boxrule=0.8pt,
    arc=1.5mm,
    auto outer arc,
    boxsep=5pt,
    left=6pt,
    right=6pt,
    top=6pt,
    bottom=6pt,
    fonttitle=\bfseries,
    before skip=10pt,
    after skip=10pt,
  }
}
\def\BibTeX{{\rm B\kern-.05em{\sc i\kern-.025em b}\kern-.08em
    T\kern-.1667em\lower.7ex\hbox{E}\kern-.125emX}}
\begin{document}

\newcommand{\authblk}[4]{%
  \begin{tabular}[t]{@{}c@{}}%
    \large{#1}\\%
    #2\\%
    #3\\%
    #4%
  \end{tabular}%
}

\title{Exploring Membership Inference Vulnerabilities in Clinical Large Language Models}


\author{%
  \makebox[\textwidth][c]{%
    \begin{tabular}{@{}c@{\hspace{2em}}c@{\hspace{2em}}c@{}}
      \authblk{Alexander Nemecek}{Case Western Reserve University}{Cleveland, Ohio, USA}{ajn98@case.edu} &
      \authblk{Zebin Yun}{Tel Aviv University}{Tel Aviv, Israel}{zebinyun@mail.tau.ac.il} &
      \authblk{Zahra Rahmani}{Case Western Reserve University}{Cleveland, Ohio, USA}{zxr81@case.edu}
    \end{tabular}%
  }\\[1.0em]%
  \makebox[\textwidth][c]{%
    \begin{tabular}{@{}c@{\hspace{0em}}c@{\hspace{0em}}c@{\hspace{0em}}c@{}}
      \authblk{Yaniv Harel}{Tel Aviv University}{Tel Aviv, Israel}{yaniv10@tauex.tau.ac.il} &
      \authblk{Vipin Chaudhary}{Case Western Reserve University}{Cleveland, Ohio, USA}{vxc204@case.edu} &
      \authblk{Mahmood Sharif}{Tel Aviv University}{Tel Aviv, Israel}{mahmoods@tauex.tau.ac.il} &
      \authblk{Erman Ayday}{Case Western Reserve University}{Cleveland, Ohio, USA}{exa208@case.edu}
    \end{tabular}%
  }%
}

\maketitle

\begin{abstract}
As large language models (LLMs) become progressively more embedded in clinical decision-support, documentation, and patient-information systems, ensuring their privacy and trustworthiness has emerged as an imperative challenge for the healthcare sector. Fine-tuning LLMs on sensitive electronic health record (EHR) data improves domain alignment but also raises the risk of exposing patient information through model behaviors. In this work-in-progress, we present an exploratory empirical study on membership inference vulnerabilities in clinical LLMs, focusing on whether adversaries can infer if specific patient records were used during model training. Using a state-of-the-art clinical question-answering model, \texttt{Llemr}, we evaluate both canonical loss-based attacks and a domain-motivated paraphrasing-based perturbation strategy that more realistically reflects clinical adversarial conditions. Our preliminary findings reveal limited but measurable membership leakage, suggesting that current clinical LLMs provide partial resistance yet remain susceptible to subtle privacy risks that could undermine trust in clinical AI adoption. These results motivate continued development of context-aware, domain-specific privacy evaluations and defenses such as differential privacy fine-tuning and paraphrase-aware training, to strengthen the security and trustworthiness of healthcare AI systems.
\end{abstract}


\section{Introduction}\label{intro}

Large language models (LLMs) have become essential components across a broad spectrum of modern computational systems~\cite{yang2024harnessing}. Their ability to generate coherent text, perform complex reasoning, and adapt across diverse tasks through instruction tuning has led to rapid adoption in domains ranging from education and law to software engineering and scientific research~\cite{wei2021finetuned, al2024analysis, rasnayaka2024empirical}. Advances in model architecture and fine-tuning strategies have further improved their contextual understanding and reasoning abilities, enabling LLMs to interpret in text and structured data, and to engage in domain-specific dialogue~\cite{wei2022chain, jiang2023structgpt}. As a result, research communities and institutions have integrated LLMs into workflows that demand high reliability, interpretability, and domain awareness~\cite{Bommasani2021FoundationModels}.

As LLMs continue to expand into high-stakes domains such as finance, defense, and healthcare, the latter has emerged as one of the most promising yet vulnerable areas of adoption~\cite{li2023large, ferrag2024generative, yang2023large}. Within the healthcare sector, these models are being further integrated into clinical question-answering systems, medical documentation tools, diagnostic assistance frameworks, and decision-support pipelines~\cite{thirunavukarasu2023large}. Their ability to process large volumes of unstructured medical text and integrate diverse data sources positions them as powerful tools for advancing precision medicine and clinical research. To achieve domain alignment and improve performance on specialized medical tasks, researchers have turned to fine-tuning LLMs on sensitive electronic health record (EHR) data by further training pre-trained models~\cite{wang2023clinicalgpt, chen2023meditron}. For instance, \texttt{Llemr}, which extends general-purpose models through training on clinical event sequences and question-answer datasets, represents a fine-tuned and architecture-adapted clinical language model purpose-built for EHR-based reasoning~\cite{wu2024instruction}. Such integrations illustrate the potential of LLMs to accelerate innovation in healthcare delivery and biomedical discovery.

Despite the transformative potential of LLMs in healthcare applications, their integration into clinical workflows introduces heightened privacy and security challenges~\cite{yu2025large}. Healthcare data represent some of the most sensitive forms of personal information, and the large-scale use of EHRs for fine-tuning amplifies the risk of inadvertent data exposure~\cite{chen2024generative, shoghli2024balancing}. These risks are compounded by the opaque nature of large models and the difficulty of controlling what information they may memorize or reveal during inference. Potential threats include unintentional data leakage through generated outputs~\cite{akkus2025generated}, model inversion attacks that attempt to reconstruct patient-specific information~\cite{fredrikson2015model}, and membership inference attacks (MIAs) that aim to determine whether a particular record was used during training~\cite{Shokri2017Membership}. Such vulnerabilities highlight the tensions between advancing medical AI capabilities and preserving patient privacy. Moreover, this emphasizes the need for systematic security evaluations of domains-specific LLMs.

From the multiple threats facing LLMs, MIAs have become one of the most direct threat models specifically for further fine-tuned models~\cite{jagannatha2021membership}. In an MIA, an adversary aims to determine whether a particular data record was included in a model's training set~\cite{Shokri2017Membership}. By observing differences in model behavior, such as output probabilities, confidence scores, or response stability, between data points that were seen during training (members) and those that were not (non-members), the attacker can infer potential memberships~\cite{truex2019demystifying}. In the context of LLMs in healthcare, a successful attack could reveal that a patient's record was used during model development, thereby exposing sensitive health information and undermining the trust needed for clinical AI adoption. 

Previous studies have investigated MIAs against general-purpose LLMs, typically using coarse perturbation strategies or loss-based signals to distinguish between member and non-member samples~\cite{he2025towards}. However, these approaches often rely on random masking, uniform noise, or generalized text perturbations that fail to capture the structured and domain-specific nature of medical data~\cite{galli2024noisy}. In particular, medical text exhibits linguistic consistency, terminology constraints, and semantic dependencies that differ from general-purpose corpora. This motivates the need for membership inference approaches that can reflect these domain characteristics through more realistic perturbation strategies~\cite{mireshghallah2022quantifying}.

To address these limitations, this work-in-progress study presents an initial investigation into the privacy vulnerabilities of clinical LLMs through the lens of MIAs. Specifically, we evaluate whether fine-tuned clinical questions-answering models trained on EHR data exhibit measurable differences in behavior between training and non-training examples. We introduce a domain-motivated, paraphrase-based perturbation strategy that simulates more realistic adversarial conditions, an early step toward context-aware membership inference methods where attackers may have access to semantically similar but not identical patient-related queries. Our preliminary experiments on a state-of-the-art, clinical LLM (\texttt{Llemr}~\cite{wu2024instruction}) demonstrate non-trivial yet bounded membership leakage, suggesting that while existing models exhibit some resilience, subtle privacy risks remain. These findings highlight the need for systematic, domain-specific evaluations of privacy and trustworthiness in medical language models. Rather than a complete framework, our goal is to establish an empirical baseline and methodological foundation for studying how linguistic and semantic variability influence privacy leakage in clinical LLMs.

In summary, our contribution is twofold: (i) we present an empirical study of membership inference in a clinical LLM fine-tuned on EHR data, and (ii) we introduce a paraphrase-based attack strategy that models realistic adversarial access in medical contexts.

The remainder of this paper is organized as follows. Section~\ref{related_work} reviews prior work on membership inference and its applications to LLMs. Section~\ref{exprimental_design} describes our experimental setup, attack models, and evaluation metrics. Sections~\ref{results} and~\ref{discussion} present and discuss the empirical results respectively, while Section~\ref{future_work} outlines directions for ongoing and future research focusing on strengthening the privacy and trustworthiness of clinical language models. Finally, Section~\ref{conclusion} concludes our work.

\section{Related Work}\label{related_work}

The study of membership inference attacks (MIAs) originated from the observation that machine learning models behave differently on data seen during training versus unseen data. Shokri et al. first demonstrated that this behavioral gap can allow adversaries to infer whether a specific sample was used during training, with overfitting serving as a major, but not exclusive, contributor to such leakage~\cite{Shokri2017Membership}. Their findings were the foundation for research exploring the generalization-privacy trade-off across various model architectures.

Building on this insight, Yeom et al. offered a theoretical formalization of the relationship between overfitting and privacy risk. They proved that for classification settings with bounded loss, the membership advantage is equivalent to the generalization error, thereby establishing a bridge between empirical risk minimization and privacy exposure~\cite{Yeom2018PrivacyRisk}. This connection further motivated analysis of how related attacks, such as attribute inference, which seeks to recover hidden features of training samples, stem from the same underlying vulnerabilities in model generalization. 

As machine learning has shifted toward large-scale generative models, recent work has revisited MIAs in the context of large language models (LLMs). Studies highlight a gap in understanding privacy risks during the pre-training phase, where models ingest massive text corpora potentially containing sensitive data~\cite{Duan2024DoMembership}. Subsequent efforts have extended MIAs to fine-tuned LLMs, introducing self-prompt calibration and differentiating between reference-based and reference-free strategies that exploit token-level probabilities and prompt sensitivity~\cite{Fu2023SPV_MIA}. Beyond these sequence-level approaches, token-level analyses such as Min-$K\%$ and its normalized variant Min-$K\%++$ probe the portions of text where models exhibit the least confidence, effectively exposing fine-grained memorization patterns~\cite{zhang2025mink}. Collectively, these investigations reveal that LLMs exhibit nuanced privacy and security behaviors that depend not only on model scale but also on task formulation and prompting strategy.

In medical and clinical language models, these concerns take on particular substance. Jagannatha et al. demonstrated that models trained on clinical narratives exhibit measurable privacy leakages up to 7\% in some configurations, and that smaller or masked-language architectures tend to leak less information than larger, auto-regressive ones~\cite{Jagannatha2021ClinicalMIA}. Such findings further emphasize that privacy leakage in healthcare AI cannot be viewed solely as a theoretical risk but as a practical challenge in protecting patient data and institutional trust.

Parallel advances in perturbation-based analysis have deepened our understanding of model robustness and uncertainty, offering complementary insights into privacy. Methods such as SPUQ use controlled perturbations to quantify uncertainty by probing model sensitivity to input variations~\cite{Gao2024SPUQ}, while other frameworks explicitly disentangle meaningful distributional changes from intrinsic stochastic noise~\cite{Rauba2024QuantifyingPerturbationImpacts}. However, these approaches generally remain model-agnostic and are not tailored to domain-constrained data such as clinical narratives. These studies collectively show that small, semantically consistent modifications to inputs can expose structural weaknesses in model confidence as well as weaknesses that adversaries might exploit for inference attacks.

Building on these threads, our work introduces a preliminary, paraphrase-based framework specifically tailored for medical question-answering systems. By leveraging natural linguistic variation rather than synthetic noise, we explore how realistic paraphrases influence model confidence and membership signals. This design represents an early, domain-motivated step toward context-aware membership inference in clinical LLMs, bridging the gap between theoretical privacy formulations and practical adversarial conditions in healthcare. To empirically assess these privacy and trustworthiness considerations, we conduct a preliminary evaluation of membership inference vulnerabilities in a representative clinical language model.

\section{Experimental Design}\label{exprimental_design}

We evaluate membership inference attacks (MIAs) against \texttt{Llemr}~\cite{wu2024instruction}, an instruction-tuned clinical language model designed to reason over electronic health records (EHRs). All experiments use de-identified records derived from the publicly available MIMIC-IV dataset~\cite{johnson2023mimic}; no protected health information (PHI) was accessed, and no institutional review board (IRB) approval was required. Each patient history is represented as a sequence of structured clinical events (e.g., laboratory tests, prescriptions). These event sequences are embedded with \texttt{ClinicalBERT}~\cite{huang2019clinicalbert} and projected into the token space of the backbone LLM following the \texttt{Llemr} architecture, which is trained in two stages: schema alignment and clinical reasoning, to enable open-ended question answering.

We adopt \texttt{Llemr} because it is an open-source, instruction-tuned clinical LLM that jointly models structured EHR streams and natural language, making it a representative benchmark for studying privacy risks that arise when clinical data are used for fine-tuning. Our intent is not to benchmark multiple models or retrain variants, but rather to examine privacy risks in an existing, state-of-the-art clinical LLM that has already undergone domain-specific fine-tuning. We obtained both the pre-fine-tuned model and the corresponding training and evaluation splits and therefore did not perform any additional fine-tuning or data modification. This choice allows us to isolate membership leakage attributable to the model's existing fine-tuning process rather than artifacts introduced by additional retraining. The in-group (member) and out-group (non-member) sets are drawn directly from these official partitions, ensuring consistency with the model's intended data distribution. While Meeus et al.~\cite{meeus2025sok} note limitations of post-hoc membership constructions, our goal is to establish an empirical privacy baseline for a representative, deployed clinical model rather than to design a generalized attack pipeline requiring retraining. Future work will explore randomized fine-tuning and multi-model replication to complement this focused, domain-grounded analysis.

To construct a balanced MIA benchmark, we uniformly sample $10{,}000$ question-answer (Q\&A) pairs from the model's training split as \emph{members} and $10{,}000$ Q\&A pairs from the held-out evaluation split as \emph{non-members}. All reported metrics aggregate over these $20{,}000$ examples. We report area under the receiver operating characteristic curve (AUC)~\cite{sankararaman2009genomic} and True Positive Rate at low False Positive Rate (TPR@FPR), following prior MIA evaluations~\cite{carlini2022membership}.

\subsection{Adversary Knowledge and Threat Model}
We study the standard \emph{black-box} threat model commonly used in the MIA literature~\cite{carlini2022membership, leino2020stolen, nasr2019comprehensive, sablayrolles2019white, salem2019ml}. The adversary may query the target model with a Q\&A pair and observe the model's token-level posteriors (equivalently, the sequence negative log-likelihood or perplexity); no gradients or parameter access are assumed. Formally, for an input text $x$ (a Q\&A pair) with token set $T$ and model parameters $\theta$, we define
\[
\begin{aligned}
\mathrm{NLL}(x) &= -\frac{1}{|T|}\sum_{t\in T}\log p_\theta\bigl(y_t \mid y_{<t}, x\bigr),\\
\mathrm{PPL}(x) &= \exp\!\bigl(\mathrm{NLL}(x)\bigr).
\end{aligned}
\]
We report both negative log-likelihood (NLL) and its exponential form, perplexity (PPL), which are standard and equivalent measures of model confidence. Lower NLL (or equivalently, lower PPL) indicates higher model certainty on a given input, and such confidence differences are often leveraged to infer membership status.

\subsection{Attack Methods}
To provide a coherent and interpretable evaluation, we organize our attacks by the \emph{granularity} of the model confidence signal they exploit. Concretely, we consider (i) \textbf{sequence-level} likelihood signals that summarize model certainty over an entire response, (ii) \textbf{semantic}-aware sequence signals that test whether near-equivalent inputs preserve any membership advantage, and (iii) \textbf{token-level} statistics that isolate local uncertainty in the hardest tokens. This progression moves from coarse to fine-grained probes of memorization and allows us to test complementary hypotheses about how membership information may manifest in clinical LLMs.

\paragraph{Loss Attack (baseline)}
Sequence-level negative log-likelihood (NLL) is the canonical signal for membership inference. If the model has memorized an example, it will typically assign higher overall likelihood (lower NLL) to that exact example. Thus, the loss attack tests whether aggregate sequence confidence suffices to separate members from non-members.

For each Q\&A pair $(q,a)$, we compute the normalized negative log-likelihood $\mathrm{NLL}(q,a)$ while conditioning the model on the question tokens (i.e., focusing on the answer tokens). The canonical likelihood-based attack uses the score
\[
s_{\mathrm{loss}}(q,a) = -\mathrm{NLL}(q,a),
\]
where a higher $s_{\mathrm{loss}}$ indicates greater membership likelihood.

\paragraph{Paraphrased Loss Attack (weakened black-box)}
Real adversaries are unlikely to possess verbatim training prompts. This attack evaluates whether membership signals persist when the adversary submits \emph{semantically equivalent but lexically distinct} queries. If a model's memorization is robust to paraphrase, then paraphrase-conditioned likelihoods should still separate members from non-members.

We synthesize paraphrased Q\&A variants of each original $(q,a)$ using \texttt{ChatGPT-3.5-Turbo}~\cite{openai2023chatgpt35} (see prompt below). For each original Q\&A we generate $m\in\{1,2,3\}$ paraphrases $\{\tilde{x}_i\}_{i=1}^m$. For each paraphrase we compute $\mathrm{NLL}(\tilde{x}_i)$ and use the sample mean
\[
\overline{\mathrm{NLL}} = \frac{1}{m}\sum_{i=1}^{m}\mathrm{NLL}(\tilde{x}_i),
\]
with the final attack signal
\[
s_{\mathrm{para}}(q,a) = -\overline{\mathrm{NLL}}.
\]
Larger $s_{\mathrm{para}}$ values indicate higher membership likelihood. Paraphrase generation is performed with a fixed random seed and a single prompt template to ensure reproducibility and control semantic fidelity.

\begin{tcolorbox}[reviewbox]
\textbf{Paraphrase prompt.} ``You are a careful paraphraser. Keep medically relevant facts unchanged. Preserve meaning; do not introduce or remove medical facts; keep entities and numerical facts unchanged.''
\end{tcolorbox}

\paragraph{Min-$K\%$ and Min-$K\%++$ attack}
After probing sequence- and semantic-level confidence, we next examine whether leakage is localized to specific token positions. Sequence-level signals may obscure localized memorization concentrated in a few low-probability tokens (e.g., specific names, dates, or rare clinical terms). Min-$K\%$ and Min-$K\%++$ target these local irregularities by focusing on the worst-performing token positions, asking whether per-token uncertainty reveals membership information that aggregate metrics miss.

Min-$K\%$~\cite{zhang2025mink} focuses on the subset of tokens where the model is least confident. For an input with token-level log-probabilities $\{\log p_t\}_{t\in T}$ and a fraction $K\in(0,1]$, Min-$K\%$ computes the average of the lowest $K$-fraction of token log-probabilities and uses this statistic as the attack score. Min-$K\%++$ refines this by first standardizing each token's log-prob relative to the model's per-step distribution  before selecting the bottom $K$-fraction. Intuitively, Min-$K\%$ emphasizes the ``weakest'' tokens that may reveal memorization, while Min-$K\%++$ removes position-specific uncertainty bias, yielding a more comparable per-token signal across positions.

Together, these complementary attacks characterize whether membership leakage in clinical LLMs arises from (i) aggregate sequence confidence, (ii) robustness to semantically equivalent perturbations, or (iii) localized uncertainty in a small subset of tokens.

\section{Results}\label{results}

Our preliminary evaluation reveals several patterns in how clinical LLMs leak (or resist leaking) membership information.

\subsection{Loss and Paraphrased Loss Attack Performance}
\begin{figure}[t]
    \centering
    \includegraphics[width=1.0\linewidth]{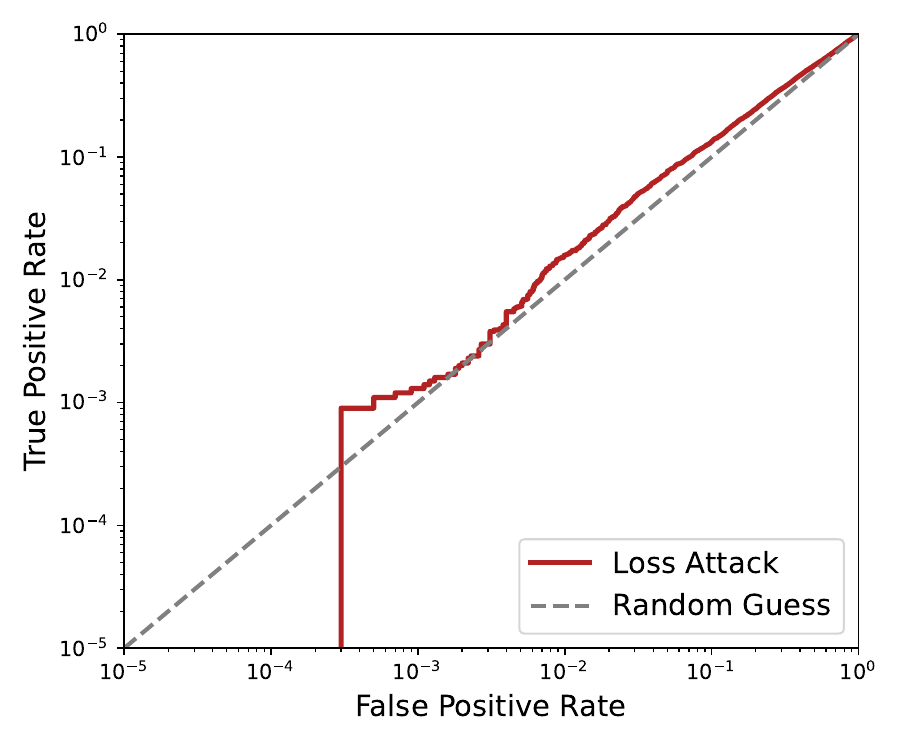}
    \caption{ROC curve for the Loss Attack. Both axes are log-scaled to highlight separability at low false positive rates. The dashed diagonal represents random guessing.}
    \label{fig:loss_attack}
\end{figure}

The baseline Loss Attack achieved an AUC of 0.5392 with TPR@1\%FPR of 1.46\%, indicating limited but non-negligible privacy leakage (Table~\ref{tab:mia-results}, Figure~\ref{fig:loss_attack}). The observed difference in average loss between members (1.962) and non-members (2.216) aligns with theoretical predictions that training examples exhibit lower loss due to mild overfitting.

\begin{figure}[t]
    \centering
    \includegraphics[width=1.0\linewidth]{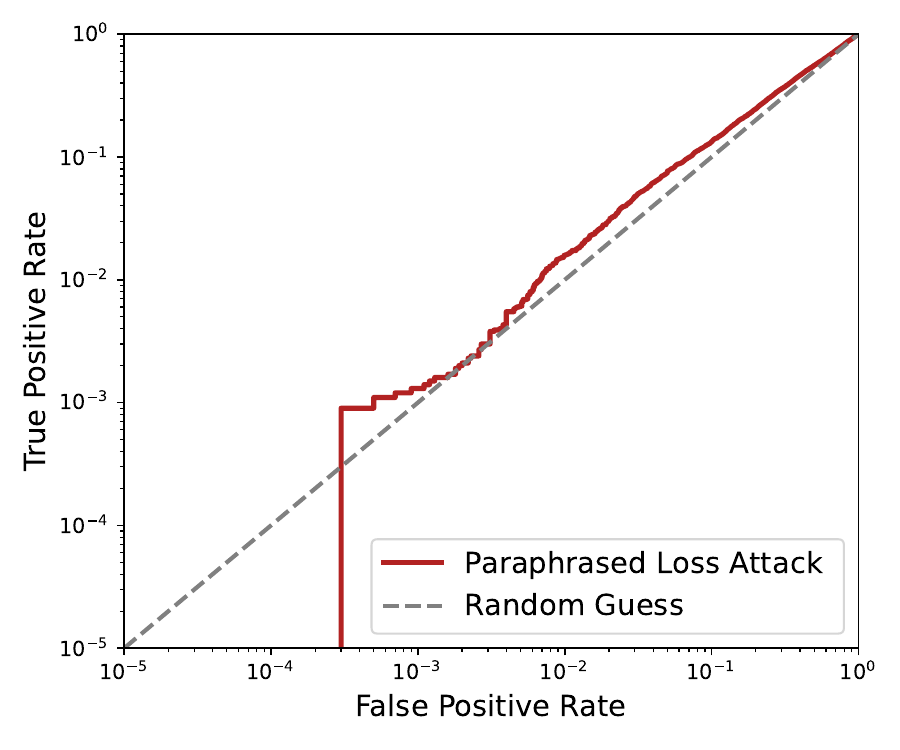}
    \caption{ROC curve for the Paraphrased Loss Attack. Both axes are log-scaled to highlight separability at low false positive rates. The dashed diagonal represents random guessing.}
    \label{fig:paraphrased_loss_attack}
\end{figure}

The Paraphrased Loss Attack represents a more realistic and practical evaluation of adversarial capability. Although it does not substantially outperform the baseline, it provides insight by addressing the unrealistic assumption that adversaries have access to the exact training samples. In realistic clinical scenarios, potential adversaries or downstream users are far more likely to query semantically similar, rather than identical, question–answer pairs. The paraphrased attack achieved an AUC of 0.5397 and TPR@1\%FPR = 1.59\% (Table~\ref{tab:mia-results}, Figure~\ref{fig:paraphrased_loss_attack}), reinforcing that small but measurable confidence differences can expose membership signals. Although the observed AUC values (0.54) are close to random guessing, such consistent separability across attacks indicates a non-zero privacy signal that warrants further exploration.

\begin{table}[h]
\centering
\caption{Membership inference attack results against \texttt{Llemr}.}
\label{tab:mia-results}
\begin{tabular}{lccc}
\toprule
\textbf{Attack} &\textbf{TPR@1\%FPR} & \textbf{TPR@10\%FPR} & \textbf{AUC} \\
\midrule
Loss &1.46\%& 13.34\% & 0.5392 \\
Paraphrased Loss & 1.59\%& 13.32\% & 0.5397 \\
\bottomrule
\end{tabular}
\end{table}

\subsection{Semantic Similarity Validation}
Beyond quantitative performance, we also validate the semantic fidelity of paraphrases used in the attack setup. To verify that the paraphrased inputs accurately emulate a realistic adversary scenario, one in which the attacker submits semantically similar but lexically distinct queries, we evaluated semantic alignment between original and paraphrased clinical Q\&A pairs. Both the original and paraphrased texts were embedded using \texttt{all-MiniLM-L6-v2}~\cite{wang-etal-2021-minilmv2}, and cosine similarity was computed between their embeddings. Mean similarity values were then calculated separately for members and non-members.

\begin{figure}[h]
\centering
\includegraphics[width=1.0\linewidth]{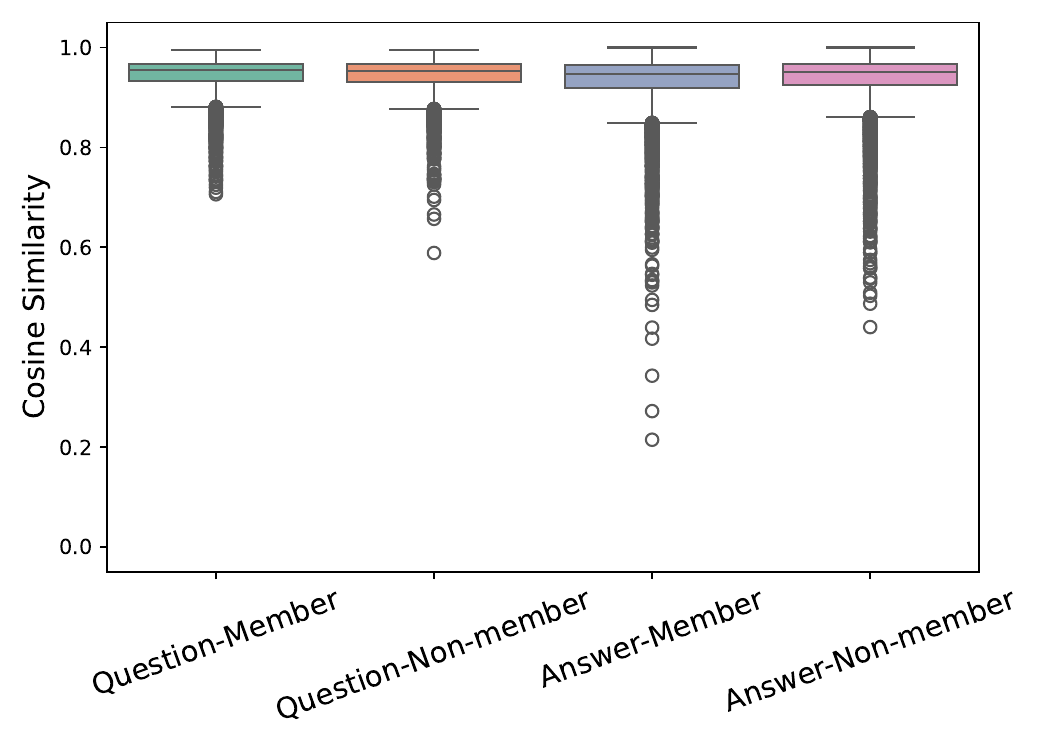}
\caption{Cosine similarity between original and paraphrased clinical Q\&A pairs.}
\label{fig:semantic_similarity_box}
\end{figure}

The paraphrases demonstrate consistently high semantic fidelity (average cosine similarity $\approx 0.93$--$0.95$), confirming that linguistic variation is introduced without altering medical meaning. For questions, members and non-members exhibit nearly identical similarity (0.947 vs. 0.945); for answers, non-members are slightly higher (0.938 vs. 0.934). These marginal differences ($\Delta\approx0.002$--$0.004$) indicate that paraphrase generation introduces realistic linguistic variation while maintaining clinical integrity.

\subsection{Min-$K\%$ Attacks} 
After examining sequence- and paraphrase-level confidence, we next probe whether membership information is localized within specific tokens rather than distributed across an entire response. Min-$K\%$ and Min-$K\%++$ address this question by focusing on the least confident token positions, capturing potential micro-level memorization that aggregate loss measures might mask.

As shown in Table~\ref{tab:mink-results}, plain Min-$K\%$ variants provide limited signal (AUC$\approx$0.51, TPR@1\%FPR$\approx$1\%), while Min-$K\%++$ yields modest improvements (AUC up to 0.5455). These results suggest that token-level normalization slightly sharpens separability but does not reveal substantially stronger membership cues. In other words, \texttt{Llemr}'s privacy exposure appears to stem primarily from global confidence patterns rather than isolated token anomalies, indicating that memorization is diffuse across the sequence rather than concentrated in rare lexical units.
\begin{table}[h]
\centering
\caption{Membership inference attack results of Min-$K\%$ and Min-$K\%++$ against \texttt{Llemr}.}
\label{tab:mink-results}
\begin{tabular}{lccc}
\toprule
\textbf{Attack} & \textbf{TPR@1\%FPR} & \textbf{TPR@10\%FPR} & \textbf{AUC} \\
\midrule
Min-$K\%$ (0.1)   & 1.07\% &11.19\%& 0.5154 \\
Min-$K\%$ (0.2)   & 0.96\% &11.39\%& 0.5139 \\
Min-$K\%$ (0.5)   & 1.03\% &11.40\% &0.5133 \\
Min-$K\%++$ (0.1) & 1.40\% &13.13\% &0.5455 \\
Min-$K\%++$ (0.2) & 0.96\% & 12.94\%&0.5139 \\
Min-$K\%++$ (0.5) & 1.37\% & 12.45\%&0.5486 \\
\bottomrule
\end{tabular}
\end{table}

\section{Discussion}\label{discussion}

Overall, the results suggest that current medical language models such as \texttt{Llemr} demonstrate partial resilience to standard membership inference techniques but are still susceptible to subtle privacy leakage. The small, yet measurable separability in model confidence underscores the importance of continued assessment, even in domain-specific models that appear robust. The paraphrased attack is particularly meaningful because it reflects a more plausible threat model. In real healthcare contexts, model users or adversaries are unlikely to recall or obtain exact training samples. Instead, they may generate semantically similar questions, mirroring real information-seeking behavior in clinical settings. Thus, evaluating models under paraphrase-based perturbations provides a stronger and more realistic test of privacy resilience. Such paraphrase-based evaluations can inform the design of defense strategies that account for naturally occurring linguistic variability among clinical users. This domain-motivated approach begins to operationalize the broader goal of developing context-aware membership inference methods that account for the structured and semantically constrained nature of clinical language.

These findings motivate exploration of privacy-preserving strategies that account for linguistic variability, such as differential privacy during fine-tuning, paraphrase-aware adversarial training, or calibration-based defenses. Moreover, they underscore the need for privacy-aware model evaluation frameworks and robust fine-tuning protocols that anticipate real-world usage patterns rather than idealized threat models.

Although the present study is limited in scale and does not include white-box or shadow-model baselines, it represents a work-in-progress effort that establishes an empirical foundation for systematic evaluation of privacy leakage in clinical LLMs. Future work will extend this analysis by varying semantic distance between paraphrases and original samples to quantify how linguistic divergence affects attack success. This will help clarify the trade-offs between linguistic flexibility and privacy protection, advancing the development of trustworthy and regulation-compliant clinical AI systems aligned with public confidence and institutional responsibility. These insights collectively frame a roadmap for future investigations into both the mechanics and mitigation of membership leakage in medical LLMs.

\section{Future Work}\label{future_work}

Building on the insights and limitations outlined in Sections~\ref{results} and~\ref{discussion}, future work will expand the experimental scope to explore broader adversarial behaviors, model characteristics, and defense strategies that influence privacy leakage in clinical language models. Our goal is to advance this preliminary framework toward a context-aware membership inference pipeline that explicitly models the structured and semantically constrained nature of clinical data.

First, we aim to expand the attack surface through patient-specific probing questions that test memorization beyond the original Q\&A pairs.
To this end, we plan to generate novel questions about the same patients, targeting different aspects such as laboratory result trends, medication durations, and treatment histories. This approach will reveal whether models memorize comprehensive patient contexts rather than just isolated question-answer mappings.

Another key direction is to investigate how model architecture and capacity affect susceptibility to membership inference. By comparing transformer-based models of varying scales, the study will quantify how model size and inductive bias influence privacy risk. Following theoretical insights on the connection between model influence and privacy leakage, future analyses will examine how the contribution of specific medical attributes correlates with attack success rates.

We also intend to explore advanced perturbation strategies beyond simple paraphrasing to encompass a range of linguistic and semantic variations.
These include (i) medical synonym substitution using domain-specific ontologies to test robustness to equivalent terminology, (ii) systematic variation of question complexity and medical terminology to assess model sensitivity to linguistic depth, (iii) multi-hop reasoning questions that require integrating information from multiple facts, and (iv) temporal perturbations that alter the timeframe of medical events. Collectively, these experiments aim to develop a more comprehensive understanding of how semantic and contextual variation affect model privacy and generalization.

In parallel, upcoming work will evaluate defensive mechanisms designed to mitigate membership leakage, including differential privacy fine-tuning and paraphrase-invariant loss regularization. Studying their trade-offs with model utility remains a crucial step toward deploying privacy-preserving clinical language models in practical healthcare environments.

Finally, subsequent studies will extend this analysis to larger and more diverse medical datasets, as well as to models operating within real clinical workflows. Such studies will ensure that empirical findings translate effectively to real-world healthcare applications where patient privacy and trust remain paramount. Together, these directions outline the next stage in operationalizing context-aware, domain-specific privacy evaluation methods for clinical LLMs.

\section{Conclusion}\label{conclusion}

In this study, we conducted a work-in-progress examination of membership inference risks in a clinical question-answering model, \texttt{Llemr}. We evaluated both a standard loss-based attack and a domain-motivated, paraphrase-based variant that simulates realistic adversarial conditions. Our findings show limited but measurable privacy leakage, with small performance separability between members and non-members, indicating that even domain-tuned medical language models are not fully immune to inference risk. These results establish an initial empirical foundation and methodological baseline for assessing privacy vulnerabilities in clinical LLMs and highlight the importance of continued research into context-aware, domain-specific privacy evaluation methods and robust, privacy-conscious modeling practices in healthcare AI. Together, these efforts support the development of safer and more privacy-aware clinical language models.

\bibliographystyle{IEEEtran}
\bibliography{references}

\begin{thebibliography}{10}
\providecommand{\url}[1]{#1}
\csname url@samestyle\endcsname
\providecommand{\newblock}{\relax}
\providecommand{\bibinfo}[2]{#2}
\providecommand{\BIBentrySTDinterwordspacing}{\spaceskip=0pt\relax}
\providecommand{\BIBentryALTinterwordstretchfactor}{4}
\providecommand{\BIBentryALTinterwordspacing}{\spaceskip=\fontdimen2\font plus
\BIBentryALTinterwordstretchfactor\fontdimen3\font minus \fontdimen4\font\relax}
\providecommand{\BIBforeignlanguage}[2]{{%
\expandafter\ifx\csname l@#1\endcsname\relax
\typeout{** WARNING: IEEEtran.bst: No hyphenation pattern has been}%
\typeout{** loaded for the language `#1'. Using the pattern for}%
\typeout{** the default language instead.}%
\else
\language=\csname l@#1\endcsname
\fi
#2}}
\providecommand{\BIBdecl}{\relax}
\BIBdecl

\bibitem{yang2024harnessing}
J.~Yang, H.~Jin, R.~Tang, X.~Han, Q.~Feng, H.~Jiang, S.~Zhong, B.~Yin, and X.~Hu, ``Harnessing the power of llms in practice: A survey on chatgpt and beyond,'' \emph{ACM Transactions on Knowledge Discovery from Data}, vol.~18, no.~6, pp. 1--32, 2024.

\bibitem{wei2021finetuned}
J.~Wei, M.~Bosma, V.~Y. Zhao, K.~Guu, A.~W. Yu, B.~Lester, N.~Du, A.~M. Dai, and Q.~V. Le, ``Finetuned language models are zero-shot learners,'' \emph{arXiv preprint arXiv:2109.01652}, 2021.

\bibitem{al2024analysis}
S.~Al~Faraby, A.~Romadhony \emph{et~al.}, ``Analysis of llms for educational question classification and generation,'' \emph{Computers and Education: Artificial Intelligence}, vol.~7, p. 100298, 2024.

\bibitem{rasnayaka2024empirical}
S.~Rasnayaka, G.~Wang, R.~Shariffdeen, and G.~N. Iyer, ``An empirical study on usage and perceptions of llms in a software engineering project,'' in \emph{Proceedings of the 1st International Workshop on Large Language Models for Code}, 2024, pp. 111--118.

\bibitem{wei2022chain}
J.~Wei, X.~Wang, D.~Schuurmans, M.~Bosma, F.~Xia, E.~Chi, Q.~V. Le, D.~Zhou \emph{et~al.}, ``Chain-of-thought prompting elicits reasoning in large language models,'' \emph{Advances in neural information processing systems}, vol.~35, pp. 24\,824--24\,837, 2022.

\bibitem{jiang2023structgpt}
J.~Jiang, K.~Zhou, Z.~Dong, K.~Ye, W.~X. Zhao, and J.-R. Wen, ``Structgpt: A general framework for large language model to reason over structured data,'' \emph{arXiv preprint arXiv:2305.09645}, 2023.

\bibitem{Bommasani2021FoundationModels}
\BIBentryALTinterwordspacing
R.~Bommasani, D.~A. Hudson, E.~Adeli, R.~Altman, S.~Arora, S.~von Arx, M.~S. Bernstein, J.~Bohg, A.~Bosselut, E.~Brunskill, E.~Brynjolfsson, S.~Buch, D.~Card, R.~Castellon, N.~S. Chatterji, A.~S. Chen, K.~A. Creel, J.~Davis, D.~Demszky, C.~Donahue, M.~Doumbouya, E.~Durmus, S.~Ermon, J.~Etchemendy, K.~Ethayarajh, L.~Fei-Fei, C.~Finn, T.~Gale, L.~E. Gillespie, K.~Goel, N.~D. Goodman, S.~Grossman, N.~Guha, T.~Hashimoto, P.~Henderson, J.~Hewitt, D.~E. Ho, J.~Hong, K.~Hsu, J.~Huang, T.~F. Icard, S.~Jain, D.~Jurafsky, P.~Kalluri, S.~Karamcheti, G.~Keeling, F.~Khani, O.~Khattab, P.~W. Koh, M.~S. Krass, R.~Krishna, R.~Kuditipudi, A.~Kumar, F.~Ladhak, M.~Lee, T.~Lee, J.~Leskovec, I.~Levent, X.~L. Li, X.~Li, T.~Ma, A.~Malik, C.~D. Manning, S.~P. Mirchandani, E.~Mitchell, Z.~Munyikwa, S.~Nair, A.~Narayan, D.~Narayanan, B.~Newman, A.~Nie, J.~C. Niebles, H.~Nilforoshan, J.~F. Nyarko, G.~Ogut, L.~Orr, I.~Papadimitriou, J.~S. Park, C.~Piech, E.~Portelance, C.~Potts, A.~Raghunathan, R.~Reich, H.~Ren, F.~Rong, Y.~H. Roohani,
  C.~Ruiz, J.~Ryan, C.~R'e, D.~Sadigh, S.~Sagawa, K.~Santhanam, A.~Shih, K.~P. Srinivasan, A.~Tamkin, R.~Taori, A.~W. Thomas, F.~Tram{\`e}r, R.~E. Wang, W.~Wang, B.~Wu, J.~Wu, Y.~Wu, S.~M. Xie, M.~Yasunaga, J.~You, M.~A. Zaharia, M.~Zhang, T.~Zhang, X.~Zhang, Y.~Zhang, L.~Zheng, K.~Zhou, and P.~Liang, ``On the opportunities and risks of foundation models,'' \emph{ArXiv}, 2021. [Online]. Available: \url{https://crfm.stanford.edu/assets/report.pdf}
\BIBentrySTDinterwordspacing

\bibitem{li2023large}
Y.~Li, S.~Wang, H.~Ding, and H.~Chen, ``Large language models in finance: A survey,'' in \emph{Proceedings of the fourth ACM international conference on AI in finance}, 2023, pp. 374--382.

\bibitem{ferrag2024generative}
M.~A. Ferrag, F.~Alwahedi, A.~Battah, B.~Cherif, A.~Mechri, and N.~Tihanyi, ``Generative ai and large language models for cyber security: All insights you need,'' \emph{Available at SSRN 4853709}, 2024.

\bibitem{yang2023large}
R.~Yang, T.~F. Tan, W.~Lu, A.~J. Thirunavukarasu, D.~S.~W. Ting, and N.~Liu, ``Large language models in health care: Development, applications, and challenges,'' \emph{Health Care Science}, vol.~2, no.~4, pp. 255--263, 2023.

\bibitem{thirunavukarasu2023large}
A.~J. Thirunavukarasu, D.~S.~J. Ting, K.~Elangovan, L.~Gutierrez, T.~F. Tan, and D.~S.~W. Ting, ``Large language models in medicine,'' \emph{Nature medicine}, vol.~29, no.~8, pp. 1930--1940, 2023.

\bibitem{wang2023clinicalgpt}
G.~Wang, G.~Yang, Z.~Du, L.~Fan, and X.~Li, ``Clinicalgpt: large language models finetuned with diverse medical data and comprehensive evaluation,'' \emph{arXiv preprint arXiv:2306.09968}, 2023.

\bibitem{chen2023meditron}
Z.~Chen, A.~H. Cano, A.~Romanou, A.~Bonnet, K.~Matoba, F.~Salvi, M.~Pagliardini, S.~Fan, A.~K{\"o}pf, A.~Mohtashami \emph{et~al.}, ``Meditron-70b: Scaling medical pretraining for large language models,'' \emph{arXiv preprint arXiv:2311.16079}, 2023.

\bibitem{wu2024instruction}
\BIBentryALTinterwordspacing
Z.~Wu, A.~Dadu, M.~Nalls, F.~Faghri, and J.~Sun, ``Instruction tuning large language models to understand electronic health records,'' in \emph{The Thirty-eight Conference on Neural Information Processing Systems Datasets and Benchmarks Track}, 2024. [Online]. Available: \url{https://openreview.net/forum?id=Dgy5WVgPd2}
\BIBentrySTDinterwordspacing

\bibitem{yu2025large}
E.~Yu, X.~Chu, W.~Zhang, X.~Meng, Y.~Yang, X.~Ji, and C.~Wu, ``Large language models in medicine: Applications, challenges, and future directions,'' \emph{International Journal of Medical Sciences}, vol.~22, no.~11, p. 2792, 2025.

\bibitem{chen2024generative}
Y.~Chen and P.~Esmaeilzadeh, ``Generative ai in medical practice: in-depth exploration of privacy and security challenges,'' \emph{Journal of Medical Internet Research}, vol.~26, p. e53008, 2024.

\bibitem{shoghli2024balancing}
A.~Shoghli, M.~Darvish, and Y.~Sadeghian, ``Balancing innovation and privacy: ethical challenges in ai-driven healthcare,'' \emph{Journal of Reviews in Medical Sciences}, vol.~4, no.~1, pp. 1--11, 2024.

\bibitem{akkus2025generated}
A.~Akkus, M.~P. Aghdam, M.~Li, J.~Chu, M.~Backes, Y.~Zhang, and S.~Sav, ``Generated data with fake privacy: Hidden dangers of fine-tuning large language models on generated data,'' in \emph{34th USENIX Security Symposium (USENIX Security 25)}, 2025, pp. 8075--8093.

\bibitem{fredrikson2015model}
M.~Fredrikson, S.~Jha, and T.~Ristenpart, ``Model inversion attacks that exploit confidence information and basic countermeasures,'' in \emph{Proceedings of the 22nd ACM SIGSAC conference on computer and communications security}, 2015, pp. 1322--1333.

\bibitem{Shokri2017Membership}
R.~Shokri, M.~Stronati, C.~Song, and V.~Shmatikov, ``Membership inference attacks against machine learning models,'' in \emph{Proceedings of the 2017 IEEE Symposium on Security and Privacy (SP)}.\hskip 1em plus 0.5em minus 0.4em\relax IEEE, 2017, p. 3–18.

\bibitem{jagannatha2021membership}
A.~Jagannatha, B.~P.~S. Rawat, and H.~Yu, ``Membership inference attack susceptibility of clinical language models,'' \emph{arXiv preprint arXiv:2104.08305}, 2021.

\bibitem{truex2019demystifying}
S.~Truex, L.~Liu, M.~E. Gursoy, L.~Yu, and W.~Wei, ``Demystifying membership inference attacks in machine learning as a service,'' \emph{IEEE transactions on services computing}, vol.~14, no.~6, pp. 2073--2089, 2019.

\bibitem{he2025towards}
Y.~He, B.~Li, L.~Liu, Z.~Ba, W.~Dong, Y.~Li, Z.~Qin, K.~Ren, and C.~Chen, ``Towards label-only membership inference attack against pre-trained large language models,'' in \emph{USENIX Security}, 2025.

\bibitem{galli2024noisy}
F.~Galli, L.~Melis, and T.~Cucinotta, ``Noisy neighbors: Efficient membership inference attacks against llms,'' \emph{arXiv preprint arXiv:2406.16565}, 2024.

\bibitem{mireshghallah2022quantifying}
F.~Mireshghallah, K.~Goyal, A.~Uniyal, T.~Berg-Kirkpatrick, and R.~Shokri, ``Quantifying privacy risks of masked language models using membership inference attacks,'' \emph{arXiv preprint arXiv:2203.03929}, 2022.

\bibitem{Yeom2018PrivacyRisk}
S.~Yeom, I.~Giacomelli, M.~Fredrikson, and S.~Jha, ``Privacy risk in machine learning: Analyzing the connection to overfitting,'' in \emph{31st IEEE Computer Security Foundations Symposium (CSF)}.\hskip 1em plus 0.5em minus 0.4em\relax IEEE, 2018, pp. 268--282.

\bibitem{Duan2024DoMembership}
\BIBentryALTinterwordspacing
M.~Duan, A.~Suri, N.~Mireshghallah, S.~Min, W.~Shi, L.~Zettlemoyer, Y.~Tsvetkov \emph{et~al.}, ``Do membership inference attacks work on large language models?'' \emph{arXiv preprint arXiv:2402.07841}, 2024. [Online]. Available: \url{https://arxiv.org/abs/2402.07841}
\BIBentrySTDinterwordspacing

\bibitem{Fu2023SPV_MIA}
\BIBentryALTinterwordspacing
W.~Fu, H.~Wang, C.~Gao, G.~Liu, Y.~Li, and T.~Jiang, ``Practical membership inference attacks against fine-tuned large language models via self-prompt calibration,'' \emph{arXiv preprint arXiv:2311.06062}, 2023. [Online]. Available: \url{https://arxiv.org/abs/2311.06062}
\BIBentrySTDinterwordspacing

\bibitem{zhang2025mink}
\BIBentryALTinterwordspacing
J.~Zhang, J.~Sun, E.~Yeats, Y.~Ouyang, M.~Kuo, J.~Zhang, H.~F. Yang, and H.~Li, ``Min-k\%++: Improved baseline for pre-training data detection from large language models,'' in \emph{The Thirteenth International Conference on Learning Representations}, 2025. [Online]. Available: \url{https://openreview.net/forum?id=ZGkfoufDaU}
\BIBentrySTDinterwordspacing

\bibitem{Jagannatha2021ClinicalMIA}
\BIBentryALTinterwordspacing
A.~Jagannatha, B.~P.~S. Rawat, and H.~Yu, ``Membership inference attack susceptibility of clinical language models,'' \emph{arXiv preprint arXiv:2104.08305}, 2021. [Online]. Available: \url{https://arxiv.org/abs/2104.08305}
\BIBentrySTDinterwordspacing

\bibitem{Gao2024SPUQ}
\BIBentryALTinterwordspacing
X.~Gao, J.~Zhang, L.~Mouatadid, and K.~Das, ``Spuq: Perturbation-based uncertainty quantification for large language models,'' \emph{arXiv preprint arXiv:2403.02509v1}, 2024. [Online]. Available: \url{https://arxiv.org/html/2403.02509v1}
\BIBentrySTDinterwordspacing

\bibitem{Rauba2024QuantifyingPerturbationImpacts}
\BIBentryALTinterwordspacing
P.~Rauba, Q.~Wei, and M.~van~der Schaar, ``Quantifying perturbation impacts for large language models,'' \emph{arXiv preprint arXiv:2412.00868}, 2024. [Online]. Available: \url{https://arxiv.org/abs/2412.00868}
\BIBentrySTDinterwordspacing

\bibitem{johnson2023mimic}
A.~E. Johnson, L.~Bulgarelli, L.~Shen, A.~Gayles, A.~Shammout, S.~Horng, T.~J. Pollard, S.~Hao, B.~Moody, B.~Gow \emph{et~al.}, ``Mimic-iv, a freely accessible electronic health record dataset,'' \emph{Scientific data}, vol.~10, no.~1, p.~1, 2023.

\bibitem{huang2019clinicalbert}
K.~Huang, J.~Altosaar, and R.~Ranganath, ``Clinicalbert: Modeling clinical notes and predicting hospital readmission,'' \emph{arXiv preprint arXiv:1904.05342}, 2019.

\bibitem{meeus2025sok}
M.~Meeus, I.~Shilov, S.~Jain, M.~Faysse, M.~Rei, and Y.-A. de~Montjoye, ``Sok: Membership inference attacks on llms are rushing nowhere (and how to fix it),'' in \emph{2025 IEEE Conference on Secure and Trustworthy Machine Learning (SaTML)}.\hskip 1em plus 0.5em minus 0.4em\relax IEEE, 2025, pp. 385--401.

\bibitem{sankararaman2009genomic}
S.~Sankararaman, G.~Obozinski, M.~I. Jordan, and E.~Halperin, ``Genomic privacy and limits of individual detection in a pool,'' \emph{Nature genetics}, vol.~41, no.~9, pp. 965--967, 2009.

\bibitem{carlini2022membership}
N.~Carlini, S.~Chien, M.~Nasr, S.~Song, A.~Terzis, and F.~Tramer, ``Membership inference attacks from first principles,'' in \emph{2022 IEEE symposium on security and privacy (SP)}.\hskip 1em plus 0.5em minus 0.4em\relax IEEE, 2022, pp. 1897--1914.

\bibitem{leino2020stolen}
K.~Leino and M.~Fredrikson, ``Stolen memories: Leveraging model memorization for calibrated $\{$White-Box$\}$ membership inference,'' in \emph{29th USENIX security symposium (USENIX Security 20)}, 2020, pp. 1605--1622.

\bibitem{nasr2019comprehensive}
M.~Nasr, R.~Shokri, and A.~Houmansadr, ``Comprehensive privacy analysis of deep learning: Passive and active white-box inference attacks against centralized and federated learning,'' in \emph{2019 IEEE symposium on security and privacy (SP)}.\hskip 1em plus 0.5em minus 0.4em\relax IEEE, 2019, pp. 739--753.

\bibitem{sablayrolles2019white}
A.~Sablayrolles, M.~Douze, C.~Schmid, Y.~Ollivier, and H.~J{\'e}gou, ``White-box vs black-box: Bayes optimal strategies for membership inference,'' in \emph{International Conference on Machine Learning}.\hskip 1em plus 0.5em minus 0.4em\relax PMLR, 2019, pp. 5558--5567.

\bibitem{salem2019ml}
A.~Salem, Y.~Zhang, M.~Humbert, P.~Berrang, M.~Fritz, and M.~Backes, ``Ml-leaks: Model and data independent membership inference attacks and defenses on machine learning models,'' in \emph{26th Annual Network and Distributed System Security Symposium, NDSS}.\hskip 1em plus 0.5em minus 0.4em\relax The Internet Society, 2019.

\bibitem{openai2023chatgpt35}
OpenAI, ``Chatgpt-3.5: Large language model,'' \url{https://platform.openai.com/docs/models/gpt-3-5-turbo}, 2023, accessed: 2025-10-09.

\bibitem{wang-etal-2021-minilmv2}
\BIBentryALTinterwordspacing
W.~Wang, H.~Bao, S.~Huang, L.~Dong, and F.~Wei, ``{M}ini{LM}v2: Multi-head self-attention relation distillation for compressing pretrained transformers,'' in \emph{Findings of the Association for Computational Linguistics: ACL-IJCNLP 2021}, C.~Zong, F.~Xia, W.~Li, and R.~Navigli, Eds.\hskip 1em plus 0.5em minus 0.4em\relax Online: Association for Computational Linguistics, Aug. 2021, pp. 2140--2151. [Online]. Available: \url{https://aclanthology.org/2021.findings-acl.188/}
\BIBentrySTDinterwordspacing

\end{thebibliography}

\end{document}